\documentclass[aps,prd,floatfix,preprint,nofootinbib]{revtex4-1}
\usepackage{graphicx}
\usepackage{epic}
\usepackage{eepic}
\usepackage{latexsym}

\newcommand{\eq}[1]{(\ref{#1})}
\newcommand{\be}{\begin{equation}}
\newcommand{\ee}{\end{equation}}
\newcommand{\bea}{\begin{eqnarray}}
\newcommand{\eea}{\end{eqnarray}}

\newcommand{\hs}[1]{\hspace{#1 mm}}

\def\a{\alpha}

\def\d{\delta}

\def\fr{\frac}

\def\m{\mu}
\def\n{\nu}

\def\r{\rho}

\def\O{\Omega}
\def\x{\xi}

\def\o{\omega}

\let\bm=\bibitem
\def\nn{\nonumber}

\begin{document}

\title{Stress-Energy Tensor of Adiabatic Vacuum in Friedmann-Robertson-Walker Spacetimes}

\author{Ali Kaya}
\email[]{ali.kaya@boun.edu.tr}
\author{Merve Tarman}
\email[]{merve.tarman@boun.edu.tr}
\affiliation{Bo\~{g}azi\c{c}i University, Department of Physics, \\ 34342,
Bebek, \.Istanbul, Turkey}

\date{\today}

\begin{abstract}

We compute the leading order contribution to the stress-energy tensor corresponding to the modes of a  quantum scalar field propagating in a Friedmann-Robertson-Walker universe with arbitrary coupling to the scalar curvature, whose exact mode functions can be expanded as an infinite adiabatic series. While for a massive field this is a good approximation for all modes when the mass of the field $m$ is larger than the Hubble parameter $H$, for a massless field only the subhorizon modes with comoving wave-numbers larger than some fixed $k_*$ obeying $k_*/a>H$ can be analyzed in this way. As infinities coming from adiabatic zero, second and fourth order expressions  are removed by adiabatic regularization,  the leading order finite contribution to the stress-energy tensor is given by the adiabatic order six terms, which we determine explicitly. For massive and massless modes these have the magnitudes $H^6/m^2$ and $H^6a^2/k_*^2$, respectively, and higher order corrections are suppressed by additional  powers of $(H/m)^2$ and $(Ha/k_*)^2$. When the scale factor in the conformal time $\eta$  is a simple power $a(\eta)=(1/\eta)^n$, the stress-energy tensor obeys  $P=\o \rho$ with $\o=(n-2)/n$ for massive and $\o=(n-6)/(3n)$ for massless modes. In that case, the adiabaticity is eventually lost when $0<n<1$ for massive and when $0<n<3/2$  for massless fields since in time $H/m$ and $Ha/k_*$ become order one. We discuss the implications of these results for de Sitter and other cosmologically relevant spaces.  

\end{abstract}

\maketitle

\section{Introduction}

Determining vacuum energy  in quantum field theory for a given physical situation is an important problem which may lead one to deduce significant  theoretical and observational results. In the simplest case where one considers a free quantum field confined in between two parallel plates in flat space, the existence of Casimir energy is experimentally verified and it agrees with the field theory calculations. This motivates one to search for possible cosmological impacts of vacuum energy at early or late times. Especially with the discovery of the recent accelerated expansion of the universe, it is important to see whether vacuum energy can play a role in acceleration. Indeed, the cosmological constant is usually thought to be related to the vacuum energy (see e.g. \cite{w}). Of course,  the problem of fixing vacuum energy in a  cosmological setting is more complicated than determining the Casimir energy associated with parallel plates. Conceptually, the most important difference is the non-uniqueness of the vacuum in an expanding universe due to absence of Poincare symmetry. Regularization in a curved space-time is also more subtle and difficult compared to the the flat space examples since it depends on the geometry in a non-trivial way. 

Adiabatic regularization \cite{ad1,ad2,ad3} is a convenient way of obtaining  finite results out of divergent stress-energy tensor expressions  in Friedman-Robertson-Walker (FRW)  space-times. In that scheme, one subtracts mode by mode contributions to the stress-energy tensor coming from a suitably defined adiabatic basis. This is in principle similar to subtracting infinite flat space contribution to get a finite Casimir energy for parallel plates. One nice feature of adiabatic regularization is that the final stress-energy tensor is guaranteed to be conserved. Moreover, it is known to be equivalent to point-splitting regularization in FRW space-times \cite{ps1,ps2}. Adiabatic regularization may suffer from infrared divergences for massless fields \cite{ir}, so extra care is needed in such cases. However, it is possible to obtain, for example, the standard trace anomaly in a relatively simple way \cite{tr}, which gives further confidence to the method. To remove quartic, quadratic and logarithmic ultraviolet divergences that generically appear in the stress-energy tensor, it is enough to determine the adiabatic mode functions up to fourth order time derivatives (i.e. up to adiabatic order four) and the corresponding subtraction terms are determined in  \cite{ad-f}. 

To specify the vacuum state in a FRW space-time, one should fix the time dependence of the mode functions in a certain way. The most convenient choice is the so called Bunch-Davies vacuum, where one  identifies the "negative frequency" solution in the remote past with the mode function corresponding to the annihilation operator (there is however an inherent ambiguity in determining the vacuum if one considers a realistic cosmological scenario  \cite{cnr}.)  As we will discuss in the next section, sometimes the mode functions can be solved as a well defined infinite series in the adiabatic expansion scheme, i.e. instead of stopping at order four  to get an approximate adiabatic solution, which would give the adiabatic subtraction terms for regularization, one can in principle continue to obtain an exact solution in series form. The vacuum associated with such mode functions is called  the {\it adiabatic vacuum}.\footnote{One usually defines adiabatic vacuum up to a certain order by identifying the initial values of the exact mode functions with the approximate adiabatic mode functions determined to that order. The adiabatic vacuum we define here has infinite order.} Naturally, the corresponding stress-energy tensor can be calculated as a series and now adiabatic regularization requires throwing out the adiabatic zero, second and the fourth order contributions, leaving the sixth order terms as the leading order contribution to the finite stress-energy tensor. Due to presence of adiabaticity, the stress-energy tensor can be thought to be related to vacuum polarization effects rather than the particle creation ones.  

In this paper, we explicitly calculate adiabatic order six terms for the stress-energy tensor of  a scalar field propagating in a FRW space-time with arbitrary coupling to the curvature scalar. As we will discuss, this is a good approximation for a massive field if the mass is larger than the Hubble parameter. For a massless field only the contributions of modes which have sufficiently large (comoving) wavenumbers can be determined in this way. As one would expect, the sixth order expressions are complicated and thus not very illuminating. However, when the scale factor of the universe is a simple power in conformal time, the stress-energy tensor   simplifies considerably. We elaborate on possible implications of our results for cosmology. Specifically, we fix  the magnitude (and the sign) of the vacuum energy density corresponding to adiabatic vacuum in cosmologically relevant spaces, and determine  when the assumption of adibaticity is a suitable approximation. 

\section{Adiabatic Vacuum}

We consider a scalar field $\phi$ which has the following action
\be
S=-\fr{1}{2}\int\sqrt{-g}\left[(\nabla\phi)^2+(m^2+\x R)\phi^2\right].
\ee
The coupling of the scalar field to the curvature scalar is governed by the dimensionless parameter $\x$. Varying the action with respect to the metric one can determine the stress-energy-momentum as 
\be
T_{\m\n}=\nabla_\m\phi \nabla_\n\phi- \xi \nabla_\mu\nabla_\nu(\phi^2) -\fr{1}{2}g_{\m\n}\left[(\nabla\phi)^2+(m^2+\xi R)\phi^2-2\xi \nabla^2 (\phi^2)\right]+\xi \phi^2 R_{\mu\nu}.\label{emt}
\ee
As our background we take the FRW  space-time with the metric
\be
ds^2=a(\eta)^2(-d\eta^2+dx^2+dy^2+dz^2).
\ee
For later use we define the Hubble parameter in conformal time as 
\be
h=\fr{a'}{a},
\ee
where the prime denotes derivative with respect to $\eta$. The quantization of the scalar field in a FRW background is straightforward. Defining a new field $\m$ by
\be
\m=a \,\phi
\ee
and applying the standard canonical quantization procedure, one can see that
the field operator $\m$ can be decomposed in terms of the {\it time-independent}
ladder operators as 
\be
\m=\int \fr{d^3
k}{(2\pi)^{3/2}}\left[\m_k(\eta)\,e^{i\vec{k}.\vec{x}}\,a_{\vec{k}}
+\m_k(\eta)^*\,e^{-i\vec{k}.\vec{x}}\,a_{\vec{k}}^\dagger\right], 
\ee
where $\vec{k}$ is the comoving momentum variable,
$[a_{\vec{k}},a^\dagger_{\vec{k}'}]=\d(\vec{k}-\vec{k}')$ and  the mode
functions satisfy  the Wronskian condition $\m_k\m_k'^*-\m_k^*\m_k'=i$ together with 
\be
\m_k''+\left[k^2+m^2a^2+(6\x-1)\fr{a''}{a}\right]\m_k=0.\label{mf}
\ee
The ground state $|0>$ of the system can be defined by imposing 
\be
a_{\vec{k}}\,|0>=0.
\ee
Using the expression for the stress-energy-momentum tensor \eq{emt}, one can
calculate the vacuum expectation values as\footnote{As it is written in \eq{emt}, there is no ordering ambiguity in the stress-energy-momentum tensor operator.}
\bea
<0|\r|0>&=&\frac{1}{4\pi^2 a^4}\int_0^\infty \left[ 
\left|\m'_k-h\m_k\right|^2+\left(k^2+m^2a^2-6\x h^2\right)|\m_k|^2+6\x h (|\m_k|^2)'\right]k^2dk,
\label{se} \\ 
<0|P|0>&=&\frac{1}{4\pi^2 a^4}\int_0^\infty \left[ \left|\m'_k-h\m_k\right|^2-\left(\fr{k^2}{3}+m^2a^2+6\x h^2\right)|\m_k|^2+6\x h (|\m_k|^2)'\right. \nn\\
&&\hs{100} \left. -2\x(|\m_k|^2)'' \right]k^2dk.\nn
\eea
The system is now fully specified except the mode function $\m_k$ obeying \eq{mf}, which would also define the vacuum state $|0>$. 

For the adiabatic expansion, one writes $\m_k$ as 
\be
\m_k=\fr{1}{\sqrt{2\O_k}}\,e^{-i\int \O_k d\eta}. \label{admf} 
\ee
Written in this way, the mode function $\m_k$ automatically satisfies the Wronskian condition. Using \eq{mf}, $\O_k$ can be seen to obey  
\be\label{11}
\O_k^2=\left[k^2+m^2a^2\right]+(6\x-1)\fr{a''}{a}+\fr{3}{4}\fr{\O_k'^2}{\O_k^2}-\fr{1}{2}\fr{\O_k''}{\O_k}.
\ee
It is possible to solve the above equation iteratively as follows: One starts from the zeroth order solution that contains no time derivatives, i.e  $\O_k^{[0]}=\sqrt{k^2+m^2a^2}$. Using $\O_k^{[0]}$ in the right hand side of \eq{11}, one can determine a second order solution $\O_k^{[2]}$, which contains terms up to two time derivatives. Now $\O_k^{[2]}$ can be used in the right hand side and this procedure can be continued iteratively to get a series solution for $\O_k$. It is important to emphasize that in this series  expansion the number of time derivatives acts like a perturbation parameter. Assuming that the final infinite sum converges, one gets a unique function $\m_k$. To obtain the second linearly independent solution, one should start the series with the negative root  $\O_k^{[0]}=-\sqrt{k^2+m^2a^2}$.

Let us try to see when the above prescription can give a well defined $\O_k$ and thus a solution for $\m_k$.  For that let us analyze the second order solution which can be found as 
\be
\O_k^{[2]}=\sqrt{k^2+m^2a^2}\left[1+\fr{(6\x-1)a''}{2a(k^2+m^2a^2)}-\fr{m^2(a'^2+aa'')}{4(k^2+m^2a^2)^2} +\fr{5m^4a^2a'^2}{8(k^2+m^2a^2)^3}\right]. \label{s1}
\ee
For a massive field, one sees that the second order terms in the square brackets have their largest values for $k=0$, which have the magnitude $H^2/m^2$, where $H$ is the Hubble parameter with respect to the proper time
\be
H=\fr{a'}{a^2}. 
\ee
Therefore, as long as $m\gg H$, the second order terms will be much smaller than the zeroth order ones even for the $k=0$ mode (note that the corrections are more suppressed for larger $k$). By inspecting the higher order adiabatic contributions, one can see that  the magnitude of the $n$'th order adiabatic terms is equal to $(H/m)^n$ for the $k=0$ mode.  Thus, for $m\gg H$ the adiabatic expansion is trustable for all modes to determine $\m_k$. 

On the other hand, for a massless field with $m=0$, the second order solution becomes 
\be
\O_k^{[2]}=k\left[1+\fr{(6\x-1)a''}{2ak^2}\right]. \label{s2}
\ee
This time higher order adiabatic corrections are suppressed by powers of $Ha/k$. Not surprisingly, adiabatic expansion fails for modes with $k/a<H$  (i.e. for superhorizon modes). However, the expansion can still be used to determine the mode functions for $k/a\gg H$ (i.e. for subhorizon  modes). Since the vacuum is defined mode by mode for each $k$, one can use adiabatic expansion to fix $\m_k$ with large enough comoving wavenumbers with $k>k_*$ for some fixed $k_*$ obeying 
\be
k_*\gg aH.
\ee
Moreover, the momentum integrals in \eq{se} can be decomposed into two decoupled pieces corresponding to the intervals $(0,k_*)$ and $(k_*,\infty)$, where in the second interval the adiabatic expansion can safely be used. 

We thus conclude that the adiabatic vacuum is physically viable\footnote{In general one may be concerned with the existence of $\O_k^{[\infty]}$ since one actually makes an asymptotic expansion about a non-analytical point  and the convergence of the series may fail in a very short time \cite{cnr}.  This should not be an issue for very massive fields or for modes with very large wave-number.} for all modes of a massive scalar field if $m\gg H$ and it can only be imposed for the ultraviolet (UV) modes of a massless scalar obeying $k\geq k_*\gg aH$, for some fixed $k_*$. 

Since the mode functions are (uniquely) specified by the adiabatic expansion scheme, one can calculate the vacuum expectation values \eq{se} for the adiabatic vacuum. For this calculation, it is convenient to express \eq{se} in terms of $\O_k$. Using \eq{admf} one finds
\bea
<0|\r|0>&=&\frac{1}{4\pi^2 a^4}\int_0^\infty \left[ \fr{\O_k}{2}+\fr{1}{2\O_k}\left(h+\fr{\O_k'}{2\O_k} \right)^2+\fr{1}{2\O_k}(k^2+m^2a^2-6\x h^2)-3\x h \fr{\O_k'}{\O_k^2}
 \right]k^2dk, \nn\\ 
<0|P|0>&=&\frac{1}{4\pi^2 a^4}\int_0^\infty \left[  \fr{\O_k}{2}+\fr{1}{2\O_k}\left(h+\fr{\O_k'}{2\O_k} \right)^2-\fr{1}{2\O_k}\left(\fr{k^2}{3}+m^2a^2+6\x h^2\right)-3\x h \fr{\O_k'}{\O_k^2}\right.\nn\\
&&\hs{90} \left. +\x\fr{\O_k''}{\O_k^2} -2\x \fr{\O_k'^2}{\O_k^3}\right]k^2dk. \label{seo}
\eea
To obtain the leading order contribution to the stress-energy tensor one should use the sixth order solution  $\O_k=\O_k^{[6]}$ and furthermore subtract the infinite zeroth, second and
fourth order adiabatic terms which can be determined by using $\O_k=\O_k^{[4]}$ again in  \eq{seo}. It  is important to recall that in this whole procedure the number of time derivatives acts like a perturbation parameter.

This is a straightforward but a very cumbersome calculation to carry out, which we perform with the help of a computer. The final result for $\O^{[6]}$ is very complicated. However, after using $\O^{[6]}$ in \eq{seo} and performing the elementary and convergent  momentum integrals,  we obtain for the massive field the following relatively simple expression\footnote{In \cite{mt}, the stress-energy tensor for the massive field has been calculated  in a covariant way from the quantum effective action obtained by  point splitting, which must be equivalent to \eq{se-m}.}
\bea
&&\rho(m)=\fr{1}{40320m^2\pi^2a^{12}}\left[(844-3528\x) a'^6-72(173-1596\x+3780\x^2)aa'^4a''+\right.\nn\\
&&12(683-6636\x+16380\x^2)a^2a'^3a'''+9(1237-14924\x+59220\x^2-75600\x^3)a^2a'^2a''^2\nn\\
&&+9(17-168\x+420\x^2)(-16a^3 a'^2 a^{(4)}+a^4a^{(3)^2}-2a^4a''a^{(4)}+2a^4a'a^{(5)})\nn\\
&&\left. +(673-8316\x+34020\x^2-45360\x^3)(2a^3a''^3-6a^3a'a''a^{(3)})\right],\nn\\
&&P(m)=\fr{1}{40320m^2\pi^2a^{12}}\left[12(211-882\x)a'^6-8(4363-39186\x+90720\x^2)
aa'^4a''\right. \nn\\
&&+4(5819-56028\x+137340\x^2)a^2a'^3 a^{(3)}+15(2389-31108\x+107100\x^2-105840\x^3)a^2a'^2a''^2\nn\\
&&+10(-473+5628\x-21924\x^2+27216\x^3)a^3a''^3+4(617-6930\x+23940\x^2-22680\x^3)a^4a''a^{(4)}\nn\\
&&+(1601-19152\x+74340\x^2-90720\x^3)a^4a^{(3)^2}-6(17-168\x+420\x^2)a^5a^{(6)}\nn\\
&&\left.+6(-3949+44828\x-160020\x^2+166320\x^3)a^3a'a''a^{(3)}+ 4(1907-18732\x+46620\x^2) a^3a'^2a^{(4)}\right.\nn\\
&&\left. +78(17-168\x+420\x^2)a^4a'a^{(5)} \right],\label{se-m}
\eea
where the numbers in the parentheses which appear above the scale factor indicate the order of $\eta$ derivatives. 

As discussed above, for a massless field only modes with $k\geq k_*$ can be treated adiabatically.  In that case, one can calculate the {\it partial} contribution of these modes to the total stress-energy tensor by changing the limits of the $k$-integrals in \eq{se} to the range $(k_*,\infty)$. Note that this partial stress-energy tensor {\it  totally decouples} from the rest of the modes and it is self consistently {\it conserved}. For the massless case we then find 
\bea
&&\rho(k_*)=\fr{(1-6\x)^2}{256k_*^2\pi^2a^9}\left[-16a'^4a''+16aa'^3a^{(3)}+(29-108\x)aa'^2a''^2-8a^2a'^2a^{(4)}\right.\nn\\
&&\left.+6(1-4\x)a^2a''^3+a^3a^{(3)^2}-2a^3a''a^{(4)}+18(4\x-1)a^2a'a''a^{(3)}+2a^3a'a^{(5)}\right],\nn\\
&&P(k_*)=-\fr{(1-6\x)^2}{768k_*^2\pi^2a^9}\left[96a'^4a''-96aa'^3a^{(3)}-(209-540\x)aa'^2a''^2+48a^2a'^2a^{(4)}\right.\nn\\
&&+(34-120\x)a^2a''^3+3(24\x-7)a^3a^{(3)^2}+4(18\x-7)a^3a''a^{(4)}+2(89-252\x)a^2a'a''
a^{(3)}\nn\\
&&\left.-14a^3a'a^{(5)}+2a^4a^{(6)}\right]. \label{se-wm}
\eea
We check that both \eq{se-m} and \eq{se-wm}  obey the conservation equation $\rho'+3h(\rho+P)=0$, as they should, since this is guaranteed by the adiabatic regularization. For each term in these expressions the total number of time derivatives acting on the scale factors always equals six and thus the magnitudes of $\rho$ and $P$ are fixed by $H^6/m^2$  and $H^6 a^2/k_*^2$, for massive and massless fields respectively, which is expected by dimensional analysis. Note that $\rho(k_*)$ and $P(k_*)$ vanish identically for the conformally coupled scalar with $\x=1/6$.

\section{Power law expansion}

The final expressions \eq{se-m} and \eq{se-wm} are not very illuminating. Therefore, in this section we focus on power law expansion in conformal time and set
\be\label{bg}
a=\left(\fr{\eta_0}{\eta}\right)^n. 
\ee
In terms of the proper time, which is defined as $dt=a d\eta$,  the scale factor becomes
\be\label{bga}
a=\left(\fr{t}{t_0}\right)^\a,\hs{10}\a=\fr{n}{n-1}. 
\ee
Since $n=\a/(\a-1)$, the metric linearly expanding in proper time with $\a=1$ must be analyzed separately, which we study at the end of this section. 

\subsection{Massive case}

In the background \eq{bg}, the stress-energy tensor of the  massive field \eq{se-m}  becomes (recall that $H=a'/a^2$ is the Hubble parameter in proper time) 
\be\label{emn}
\rho(m)=C_n\,\fr{H^6}{m^2},\hs{10} P(m)=\,\fr{n-2}{n}\,\rho(m),
\ee
where the constant $C_n$ is given by
\bea
&&C_n=\fr{1}{20160\pi^2n^4}\left[3060-4742n-4029n^2+4716n^3+1365n^4\right.\nn\\
&&-126(240-438n-389n^2+5n^3(92+37n))\x+3780(n+1)(20-74n+25n^2+35n^3)\x^2\nn\\
&&\left. -22680n(n+1)^2(11n-10)\x^3\right]. 
\eea
From the equation of state \eq{emn},  we see that for $0<n<2$ the sign of the adiabatic vacuum pressure is the opposite of the energy density. This corresponds to the range $\a>2$ or $\a<0$ in \eq{bga}. Note that while for $\a>2$ the background is accelerating with decreasing Hubble parameter, for $\a<0$ the Hubble parameter increases and there is a big-crunch singularity at finite proper time. 

It is interesting to compare the vacuum energy density with the background energy density driving the metric \eq{bg}. From the Friedmann equation, we know that $H^2\sim \rho_B /M_p^2$, where $\rho_B$ is the background energy density and $M_p$ is the Planck mass. Thus one has  
\be
\fr{\rho(m)}{\rho_B}\sim \fr{H^4}{M_p^2m^2}. 
\ee
Since adiabaticity requires $m\gg H$, unless $H\gg M_p$ the vacuum energy density $\rho(m)$ is much smaller than the background energy density $\rho_B$ showing that the  backreaction effects can safely be ignored in this setup. 

For $0<n<1$, which corresponds to $\a<0$, the Hubble parameter $H$ increases and thus adiabaticity will eventually be lost since $H/m$ grows to become order one at some time before reaching the big-crunch singularity. Although the vacuum energy density $\rho(m)$ also increases in time, when adiabaticity is lost, i.e. when $H/m={\cal O}(1)$, its ratio to the background energy density is  given by $H^2/M_p^2$ and thus $\rho(m)$ is still negligible unless $H$ is of the order of Planck scale. In any case, it would be interesting to study if the adiabatic vacuum can help to avoid the big-crunch singularity, along the lines of \cite{bigrip}. 

\begin{figure}
\centerline{
\includegraphics[width=7cm]{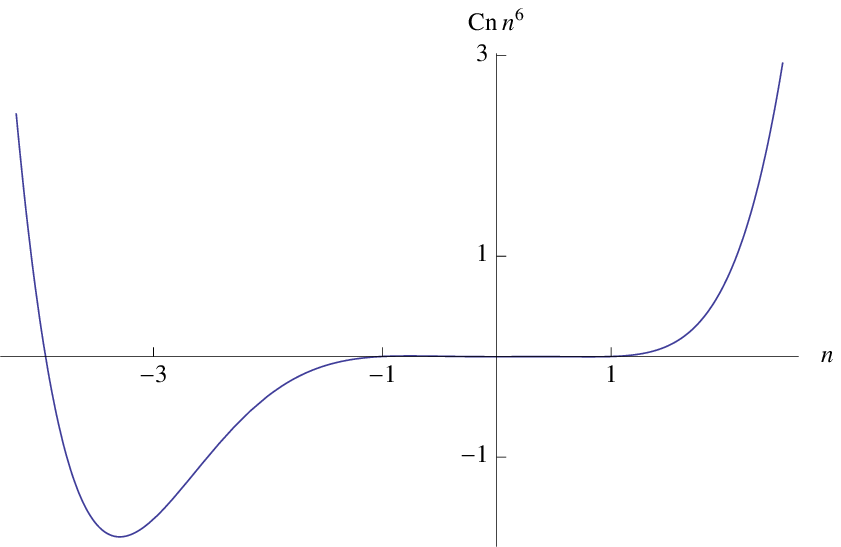} \hs{6}\includegraphics[width=7cm]{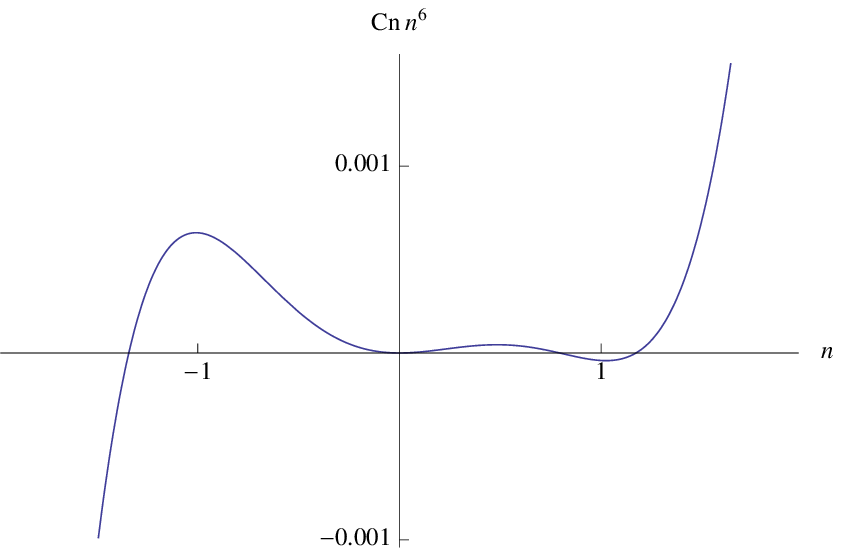}}
\caption{The graphs of $n^6C_n$ for $\x=0$ (left) and $\x=1/6$ (right). $C_n$ is multiplied by $n^6$ since $H$ contains a factor of $n$ coming from $a'$. Although it cannot be seen in the graph for $\x=0$, the curve actually oscillates in the $(-1,1)$ interval crossing the $n$-axis three times.}
\label{fig1}
\end{figure}

For minimally and conformally coupled scalars, i.e. for $\x=0$ and $\x=1/6$, the proportionality constant $C_n$ simplifies a lot:  
\be
C_n=\begin{cases}{\fr{1}{5040\pi^2n^4}\left[30-41n-15n^2+24n^3\right]\hs{10}\x=\fr{1}{6},\cr
\fr{1}{20160\pi^2n^4}\left[3060-4742n-4029n^2+4716n^3+1365n^4\right]\hs{5}\x=0.}\end{cases}
\ee
In figure \ref{fig1}, we give plots of $C_n$ for $\x=0$ and $\x=1/6$. It can be seen from the graphs that there are intervals  of the power $n$ for which $C_n$ is positive or negative. There also exists special values for which $C_n$ vanishes. Therefore, depending on the expansion power, the (leading order contribution to) the vacuum energy density can be positive, negative or even zero. 

Let us finally focus on  important special cases. In de Sitter space, which corresponds to $n=1$, 
the equation of state becomes $P(m)=-\rho(m)$ and thus adiabatic vacuum energy density is equivalent to a cosmological constant. In this case $C_n$ is given as
\be
C_1=\fr{270-7308\x+45360\x^2-90720\x^3}{20160\pi^2}.
\ee
For $\x=0$, $C_1>0$ and for $\x=1/6$, $C_1<0$, thus while the vacuum energy density is positive for the minimally coupled scalar, it is negative for the conformally coupled case, modifying the background cosmological constant correspondingly. It is interesting to compare these findings with the massive scalar field in Bunch-Davies vacuum studied in \cite{bd}. For $m\gg H$, the vacuum energy density  in Bunch-Davies vacuum is given by $\rho_{BD}\simeq -m^4$ (see \cite{cw} or eq. (3.15) of \cite{bd}). Therefore, the vacuum energy densities corresponding to Bunch-Davies and adiabatic vacua  differ both in magnitude and in sign. 

In the radiation dominated  background with $n=-1$, the vacuum pressure satisfies $P(m)=3\rho(m)$, which corresponds to a very stiff matter. In that case the constant $C_n$ becomes
\be
C_{-1}=\fr{211-882\x}{10080\pi^2}.
\ee
The vacuum energy density in the radiation dominated universe turns out to be positive both for the minimally and the conformally coupled scalars. 

Finally, in the matter dominated background one has $n=-2$ and  $P(m)=2\rho(m)$, again equivalent to a very stiff matter. The constant $C_n$ becomes
\be
C_{-2}=\fr{36(3\x-1)(29+96\x)\x-139}{10080\pi^2}.
\ee
One can see that both for $\x=0$ and for $\x=1/6$,  $C_{-2}<0$ and the vacuum energy density becomes negative. 

\subsection{Massless case}

In the background \eq{bg}, the stress-energy tensor \eq{se-wm} corresponding to the UV modes of the massless field with $k>k_*$  becomes
\be\label{emmn}
\rho(k_*)=D_n\,\fr{H^6a^2}{k_*^2},\hs{10} P(k_*)=\,\fr{n-6}{3n}\,\rho(k_*),
\ee
where the constant $D_n$ is given by
\be
D_n=\fr{1}{128\pi^2n^6}\left[5(1-6\x)^2n^2(n+1)(2-n)(2+n(n+1)(6\x-1))\right].
\ee
In that case the adiabatic vacuum pressure has the opposite sign compared to the vacuum energy  density for $0<n<6$, which is equivalent to $\a<0$ or $\a>6/5$. On the other hand, for $0<n<3/2$, which corresponds to $\a>3$ or $\a<0$, the vacuum energy density grows and adiabaticity will be lost in time since $Ha/k_*$ grows. Interestingly, in de Sitter space with $n=1$, while the  stress-energy tensor vanishes identically for the conformally and the minimally coupled scalars, the energy density grows for a generic $\x$, i.e. it is not in the form of a cosmological constant. In that case however,  the vacuum energy density still stays much smaller than the background energy density as long as the adiabatic approximation holds . 

It is important to emphasize that for the high energy modes with $k\gg a H$, the adiabatic vacuum is equivalent to Bunch-Davies vacuum, which has the following mode functions
\be\label{bdmm}
\m_k^{BD}=\eta\sqrt{\fr{k}{2}}h_u(k\eta),
\ee
where $h_u$ denotes spherical Hankel function of first kind and 
\be
u=-\fr{1}{2}+\sqrt{\fr{1}{2}+n(n+1)(1-6\x)}.
\ee
Using the asymptotic form of the spherical Hankel functions, one can  find as $k\to\infty$ that 
\be
\m_k^{BD}\to \fr{(-i)^{n+1}}{\sqrt{2k}}e^{-ik\eta}(1-6\x)\left(1-i\fr{n(n+1)}{2k\eta}-
\fr{n(n+1)\left[(1-6\x)n(n+1)-2\right]}{8k^2\eta^2}...\right),
\ee
which exactly coincides up to a phase with the adiabatic expansion (this can easily be checked for the the two terms written above). Note that for some special values, like $n=-1$, the series terminate and one ends up with a finite series instead of getting infinitely many terms. On the other hand, for infrared (IR) modes adiabatic expansion fails but \eq{bdmm} can still be used for Bunch-Davies vacuum (however the vacuum expectation values are now plagued by IR divergences, see \cite{ir}). 

\begin{figure}
\centerline{
\includegraphics[width=7cm]{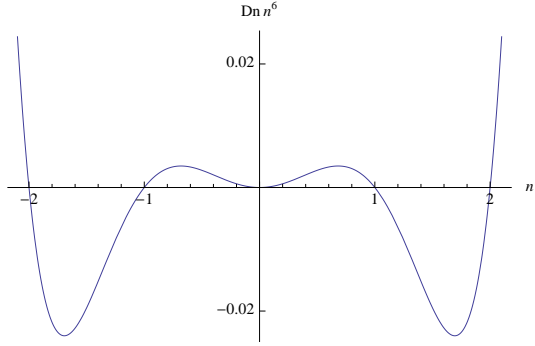}}
\caption{The graph of $n^6D_n$ for $\x=0$. $D_n$ is multiplied by $n^6$ since $H$ contains a factor of $n$ coming from $a'$.}
\label{fig2}
\end{figure}

For the minimally coupled scalar, the $D_n$ coefficient becomes
\be
D_n=\fr{5(n^4-5n^2+4)}{128\pi^2n^4}.
\ee
Curiously, the stress-energy tensor vanishes in the de Sitter space ($n=1$) and in the radiation dominated ($n=-1$) and the matter dominated ($n=-2$) universes (it also vanishes for $n=2$ corresponding to $a(t)=t^2$ in proper time).   As before, depending on $n$ the vacuum energy density can be positive or negative, which can be seen from the plot of $D_n$ given in figure \ref{fig2}.  

\subsection{The linear expansion $\a=1$}

In the linearly expanding universe where $a(t)=t/t_0$, the scale factor in the conformal time becomes
\be
a=\exp\left[\fr{\eta}{\eta_0}\right]. 
\ee
Using \eq{se-m}, one can see that the stress-energy tensor of the massive scalar obeys $P(m)=\rho(m)$ where
\be
\rho(m)=\fr{(1-6\x)^2(13-66\x)}{192m^2\pi^2}H^6.
\ee
On the other hand, the stress-energy tensor for the massless UV modes satisfies $P(k_*)=\rho(k_*)/3$, where
\be
\rho(k_*)=\fr{5(1-6\x)^3}{128k_*^2\pi^2}\, a^2\,H^6.
\ee
In both cases, the stress-energy tensor vanishes for the conformally coupled scalar and it is positive for the minimally coupled one.  Note that the equation of state can be obtained from \eq{emn} and \eq{emmn} by taking $n\to\infty$ limit, which corresponds to $\a=1$. 

\section{Conclusions}

In this paper, we determine the leading order contribution of the adiabatic mode functions to the vacuum expectation value of the stress-energy tensor by explicitly calculating adiabatic order six terms. Due to the assumption of adiabaticity, the particle creation effects are absent in this setup and thus the resulting expressions can be thought to be related to vacuum polarization effects. 
When the scale factor of the universe is given by a simple power,  the stress-energy tensor simplifies a lot and it obeys a simple equation of state. In such a background,  the vacuum energy density is proportional to a single term, which can actually be fixed by dimensional analysis. We determine when it is possible to use adiabatic approximation and when it fails in time. Depending on the power of the scale factor, (the leading order contribution to) the vacuum energy density can be  positive, negative or zero. Moreover,  it can be observed to increase, decrease or stay constant for different values of the power. 

For a massive field it is natural to assume adiabatic vacuum  when  $m\gg H$. For example 
at early times, the stabilized moduli, which are known to exist in string theory motivated models,  must satisfy this condition (this is usually required for not to change the standard cosmological picture, which is known to be in good agreement with observations). At late times, nearly all known massive fields  (neutrinos can be an exception)  obey this condition  and thus assuming adiabatic vacuum for them is reasonable. For massless fields, only the high energy modes can be thought to evolve adiabatically, where the critical scale is determined by the Hubble parameter. Moreover, the adiabatic vacuum coincides with the Bunch-Davies vacuum for UV modes,  at least when the scale factor is a simple power (although we are not aware of any proof, this is plausibly true for any given scale factor). 

We see that in the adiabatic vacuum, the energy densities corresponding to a very massive scalar field and the UV modes of a massless scalar field  turn out to be very small compared to the background energy density driving the expansion (As noted earlier  that this is true even when the leading order contribution to the vacuum energy density grows in time and adiabaticity is eventually lost.  However, it would be interesting to study the evolution of the vacuum energy density in such a case.) This supports the idea that the cosmological constant problem is an IR issue rather than being a UV problem, which is contrary to the common thought emphasizing  the huge order of magnitude discrepancy between the expected and observed values of the cosmological constant. Namely, with a proper regularization the contributions of the high energy modes to the vacuum energy density become negligible. Physically the situation is very similar to the Casimir energy associated with parallel plates where the vacuum energy density turns out to be fixed by the distance between plates, i.e. by the IR scale rather than by the UV scale (moreover it turns out for Casimir energy that the massive fields play no role and can safely be ignored).  Since experimentally measured  Casimir energy is in good agreement with  theoretical calculations, the previous statement  can be claimed to have a sound basis. Of course, the IR problem can be seen to be more challenging, but in any case it is important to pin down the real issue and our findings support the idea that  IR physics plays the key role in the cosmological constant problem.

\end{document}